\documentclass[aps,prl,twocolumn,showpacs,superscriptaddress,groupedaddress]{revtex4}
\usepackage{graphicx}   
\usepackage{dcolumn}   
\usepackage{bm}             
\usepackage{amssymb} 
\usepackage{graphicx}
\usepackage{subfigure}
\usepackage{epstopdf}
\usepackage{epsfig}
\usepackage{natbib}
\input{Qcircuit}
\begin{document}
\bibliographystyle{plainnat}
\title{Quantum Simulation of Tunneling in Small Systems}
\author{Andrew T. Sornborger}
\affiliation{Department of Mathematics and Faculty of Engineering, University of Georgia, Athens, Georgia 30602, USA}
\begin{abstract}
\noindent
A number of quantum algorithms have been performed on small quantum computers; these include Shor's prime factorization algorithm, error correction, Grover's search algorithm and a number of analog and digital quantum simulations. Because of the number of gates and qubits necessary, however, digital quantum particle simulations remain untested. A contributing factor to the system size required is the number of ancillary qubits needed to implement matrix exponentials of the potential operator. Here, we show that a set of tunneling problems may be investigated with no ancillary qubits and a cost of one single-qubit operator per time step for the potential evolution. We show that physically interesting simulations of tunneling using 2 qubits (i.e. on 4 lattice point grids) may be performed with 40 single and two-qubit gates. Approximately $70$ to $140$ gates are needed to see interesting tunneling dynamics in three-qubit (8 lattice point) simulations.
\end{abstract}
\maketitle

\noindent
Quantum simulations on quantum computers are one of a set of algorithms that give exponential improvement in computational resources relative to the best classical algorithm. Small quantum simulations have already been realized on NMR \citep{TsengEtAl1999,SomarooEtAl1999,KhitrinFung2001,NegrevergneEtAl2005,PengEtAl2005,BrownEtAl2006,PengEtAl2009,DuEtAl2010}, atomic \citep{EdwardsEtAl2010,KinoshitaEtAl2004}, ion trap \citep{FriedenauerEtAl2008,GerritsmaEtAl2010,GerritsmaEtAl2011,LanyonEtAl2011,LanyonEtAl2010} and photonic \citep{MaEtAl2011,KassalEtAl2011} quantum computers in essentially two forms: 1) analog simulations in which a quantum Hamiltonian (typically many-body or multiple spin) is mapped either directly or via a suitable pulse-sequence to a computational Hamiltonian, and 2) digital simulations in which a quantum system's Hamiltonian is split into free and interacting operators, then, using Trotter's formula, is simulated on a quantum computer. Digital quantum particle simulations \citep{Zalka1998} such as those proposed for the simulation of chemical dynamics \citep{KassalEtAl2008} have remained untested due to the large number of gates and/or ancillary qubits needed to compute the kinetic and potential operators \citep{KassalEtAl2008,BenentiStrini2008}.

{\it Digital Quantum Particle Simulations} The standard digital quantum simulation algorithm for a particle on a one-dimensional grid \citep{Zalka1998,BenentiStrini2008} encodes the position efficiently in $n = \log_2{N}$ qubits, where $N$ is the size of the lattice of discretized particle locations, $x_k = k \Delta x$, $k = 0, \dots, N-1$. The method uses a split operator approach to integrate a Schr\"odinger equation with a time-independent Hamiltonian that is first-order accurate in the time step, $\Delta t$ \citep{Zalka1998}:
\begin{eqnarray}
   | \psi(t) \rangle & = & e^{-i H t} | \psi_{init} \rangle \nonumber \\
   & = & e^{-i (V + K) t} | \psi_{init} \rangle \nonumber \\
   & = & (e^{-i V \Delta t} e^{-i K \Delta t}e^{O(\Delta t^2)})^\frac{t}{\Delta t} | \psi_{init} \rangle \;\; . \nonumber
\end{eqnarray}
Higher order methods that give more accurate time integration have been developed \citep{Yoshida1990,SornborgerStewart1999,HatanoSuzuki2005}, but methods of order higher than two require more gates per time step. Therefore we will only consider first-order methods here. States in the qubit Hilbert space
\begin{equation} 
| \psi \rangle = | \dots \psi_2 \psi_1 \psi_0 \rangle = \dots |\psi_2 \rangle |\psi_1 \rangle |\psi_0 \rangle \;\; , \nonumber
\end{equation}
where $| \psi_i \rangle \in \{ | 0 \rangle, | 1 \rangle \}$, represent particle location in the binary representation, $|x \rangle = \sum_{j\in 0,\dots,n-1} 2^j \psi_j |\psi \rangle$. The matrix exponential for the kinetic operator is calculated using a quantum Fourier transform (QFT)
\begin{equation} \nonumber
   e^{-i K t} = F e^{-i T t} F^\dagger \;\; ,
\end{equation}
where $F^\dagger$ is a discrete Fourier transform operator and $T$ is diagonal with entries proportional to $-q^2/2m$, and $q$ denotes the Fourier mode wavenumber. The resulting periodic, shift invariant unitary transform, $e^{-i K \Delta t}$, gives an approximation to the time evolution on the lattice due to the kinetic energy operator that is accurate to $N$'th order in space. This leads to the digital quantum particle simulation algorithm:
\begin{equation} \nonumber
   | \psi(t) \rangle := (e^{-i V \Delta t} F e^{-i T \Delta t} F^\dagger)^\frac{t}{\Delta t} | \psi(0) \rangle \;\; .
\end{equation}
The QFT takes of order $n^2$ gates to calculate \citep{Coppersmith1994} and general algorithms implementing the diagonal $T$ and $V$ operators require ancillary qubits \citep{Childs2004}, although it has been shown that the quadratic kinetic energy operator may be computed with $n^2$ two-qubit gates with no ancillary qubits \citep{BenentiStrini2008}. Thus, the qubit count for this algorithm is dominated by the calculation of the diagonal potential operator.

{\it Square-well Potentials} A special case of considerable interest is that of the square well potential. Square-well potentials are commonly used in the study of a number of quantum tunneling phenomena. A set of square-well potentials may be implemented with a sole single-qubit operator and no ancillary qubits. To see this, consider the single-qubit $Z$-rotation on the highest order qubit
\begin{equation} \nonumber
e^{-i V \Delta t} = e^{-i v \sigma^{n-1}_z \Delta t} = e^{-i v \sigma_z \Delta t} \otimes I \otimes I \dots \;\; ,
\end{equation}
where $v$ is a parameter, a superscript indicates the qubit to which the operator is applied and $\sigma_z$ is the Pauli $z$-matrix
\begin{equation} \nonumber
   \sigma_z = \left( \begin{array}{cc} 1 & 0 \\ 0 & -1 \end{array} \right) \;\; .
\end{equation}
The operator $e^{-i V \Delta t}$, when acting on a lattice state, implements a square-well potential by rotating qubit states with $|0 \rangle$ ($|1\rangle$ resp.) highest order qubit with positive (negative resp.)  phase velocity $v$. The single-qubit operator acting on the next highest order qubit
\begin{equation} \nonumber
e^{-i V \Delta t} = e^{-i v \sigma^{n-2}_z \Delta t} = I \otimes e^{-i v \sigma_z \Delta t} \otimes I \dots
\end{equation}
implements a double-square-well potential, and so on, with the last potential in this series implementing a comb-like potential.

\begin{figure}
\includegraphics[width = 3.5in, height = 1.3in]{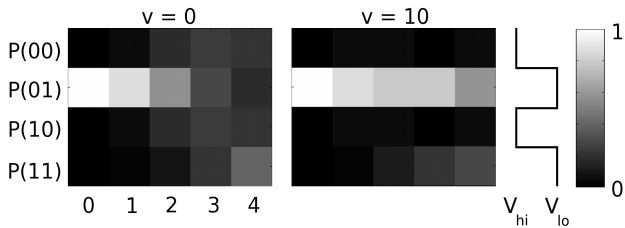}
\caption{Particle probability distributions as a function of time for the first four steps of a two qubit simulation for $v = 0$ (free particle) and $v = 10$ (particle in double well). The double-well potential and gray scale used to plot the probabilities are depicted to the right.
\label{twoqubit}}
\end{figure}

{\it A Two Qubit Tunneling Simulation} Let us consider the smallest possible tunneling simulation. An $n=2$ qubit simulation of $N = 4$ lattice points may be performed with a double well potential, $\exp(-i V \Delta t) = \exp(-i v \sigma^0_z \Delta t)$. The QFT may be computed with the operators \citep{Coppersmith1994}
\begin{equation} \nonumber
   F^\dagger = H_0 \Omega_{01} H_1\; ,
\end{equation}
where $H_i$ is a Hadamard operator on the $i$'th qubit and the controlled-phase gate $\Omega_{01} = \mathrm{diag}(1, 1, 1, \omega)$
with $\omega = \exp(2 \pi i/4)$. Note that $F^\dagger$ results in a bit-swapped Fourier transform (in our notation, $F$ is the {\it inverse} Fourier transform matrix).

The kinetic operator is then $K = F D F^\dagger$, with $D = \exp(-i \; (-2\pi/4)^2 \; \mathrm{diag}(0, 4, 1, 1) \Delta t)$ (Note that this operator is also bit-swapped and we have taken $m = 1/2$). This operation may be achieved up to an overall phase with the operator
\begin{equation} \nonumber
D = \Phi_{01}Z_1Z_0
\end{equation}
where the single-qubit operators
\begin{eqnarray}  
  Z_0 & = & e^{-i \gamma c_0 \sigma^0_z \Delta t} \nonumber \\
  Z_1 & = & e^{-i \gamma c_1 \sigma^1_z \Delta t} \nonumber
\end{eqnarray}
and $\Phi_{01}$ is a controlled-phase operator on qubits $0$ and $1$,
\begin{equation} \nonumber
  \Phi_{01} = e^{-i \gamma c_2 \; \mathrm{diag}(1,1,1,-1)_{01} \Delta t} \; ,
\end{equation}
where $\gamma = (-2\pi/4)^2/\sqrt{4}$. The coefficients ($c_0 = -1$, $c_1 = -4$ and $c_2 = 4$) in the unitary operators $Z_0$, $Z_1$ and $\Phi_{01}$ (resp.) were obtained by noting that the vectors $(1, 1, -1, -1)$, $(1, -1, 1, -1)$ and $(1, 1, 1, -1)$ that form their diagonal elements are a basis for zero mean vectors (i.e. neglecting the constant phase proportional to $(1,1,1,1)$) for $\mathbb{R}^4$. 

\begin{small}
\begin{widetext}
\begin{center}
\mbox{
\Qcircuit @C=.8em @R = .5em {
& |\psi_0 \rangle &&& \qw              & \ctrl{1}                                           & \qw & \gate{H_0} & \gate{Z_0} & \qw & \ctrl{1} & \gate{H_0} & \qw & \ctrl{1}                                           & \qw              & \gate{P_0} & \qw\\
& |\psi_1 \rangle &&& \gate{H_1} & \gate{\Omega_{01}}  & \qw & \qw              & \qw             & \gate{Z_1}             & \gate{\Phi_{01}}             &  \qw              & \qw & \gate{\Omega_{01}^\dagger} & \gate{H_1} & \qw & \qw
}
}\\
\vskip .2cm
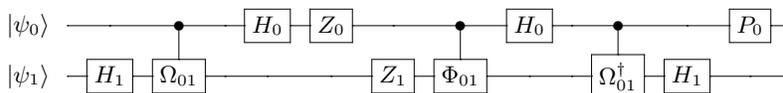
One Time Step of a Two-qubit Digital Quantum Single-particle Simulation Circuit
\end{center}
\end{widetext}
\end{small}

This method of forming a diagonal Hamiltonian using a basis of diagonal operators is not, in general, efficient and scales as $2^n-1$. However, for small systems of qubits it takes fewer gates. For instance, the Benenti-Strini (BS) algorithm \citep{BenentiStrini2008} for forming a quadratic function on the diagonal requires $n^2$ two-qubit operators. Note, however, that for $n = 2,3$ and $4$, we would need $4,9$ and $16$ (resp.) operators for BS, but at most $3,7$ and $15$ (resp.) are required using a basis along the diagonal. Furthermore, fewer multi-qubit operations are necessary with this method, although a three-qubit operator may be necessary for three-qubit simulations and three- and four-qubit operators for four-qubit simulations. Here, we use the method to construct the diagonal operator $D$ for the kinetic energy operator, however, arbitrary potentials could also be constructed this way. Because our goal is to find interesting simulations with few gates, we do not pursue more complex potentials here.

In total, each time step in this tunneling simulation requires $10$ operations, $7$ single qubit operations and $3$ two-qubit operations. The circuit for a single time step is shown above. This circuit implements a double-well potential with the gate $P_0 = \exp(-i v \sigma^0_z \Delta t)$ acting on the lowest-order qubit.

In Fig. 1, we plot lattice occupation probabilities from two simulations with a double-well potential  with $\Delta t = 1/10$: one a free-particle simulation with $v = 0$ and the other a tunneling simulation with $v = 10$. The initial state was $|\psi_{init} \rangle = |01\rangle$, corresponding to a particle in one of the wells. The free-particle probability distribution spreads across all lattice points as it evolves, whereas the particle tunnels from the well at lattice point $1$ ($|01\rangle$) to the well at lattice point $3$ ($|11\rangle$) in the tunneling simulation. These results show that differences in the evolution of the probability distribution are evident within $4$ time steps. Thus, such a simulation may be implemented on a quantum computer with $4 \times 10 = 40$ gates ($28$ single-qubit and $12$ two-qubit).

{\it Multi-Qubit Quantum Tunneling Simulations} Larger simulations require more gates. For instance, a three-qubit simulation requires $6$ gates per QFT ($3$ single-qubit and $3$ two-qubit), $6$ gates for the diagonal kinetic energy operator ($3$ single-qubit and $3$ two-qubit), and one single-qubit gate for the potential as shown in the circuit diagram below.
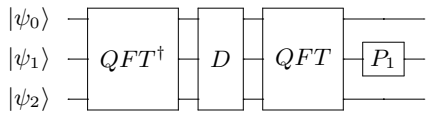
\begin{small}
\begin{center}
\mbox{
\Qcircuit @C=.8em @R = .5em {
& |\psi_0 \rangle &&& \multigate{2}{QFT^\dagger}  & \multigate{2}{D}  & \multigate{2}{QFT} & \qw & \qw \\
& |\psi_1 \rangle &&& \ghost{QFT^\dagger}              & \ghost{D}             & \ghost{QFT}              & \gate{P_1} & \qw \\
& |\psi_2 \rangle &&& \ghost{QFT^\dagger}              & \ghost{D}             & \ghost{QFT}              & \qw    & \qw         
}
}
\end{center}
\end{small}
where $P_1 = \exp(-i v \sigma^1_z \Delta t)$ is shown, representatively, for a double-well potential, but other square-well potentials could be generated by acting on different qubits,
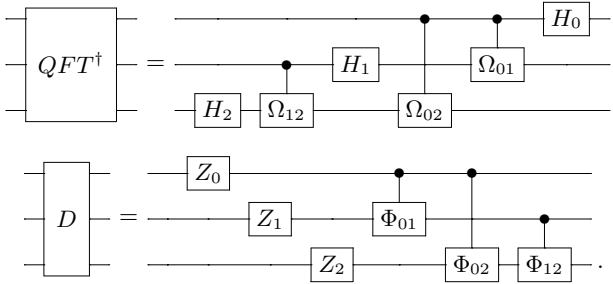
\begin{small}
\begin{center}
\mbox{
\Qcircuit @C=.8em @R = .5em {
& \multigate{2}{QFT^\dagger} & \qw &    & & \qw              & \qw                                & \qw              & \ctrl{2}                          & \ctrl{1}                         & \gate{H_0} & \qw\\
& \ghost{QFT^\dagger}            & \qw & = & & \qw              & \ctrl{1}                           & \gate{H_1} &  \qw                              & \gate{\Omega_{01}} & \qw       & \qw       \\
& \ghost{QFT^\dagger}            & \qw &    & & \gate{H_2} & \gate{\Omega_{12}}  & \qw               & \gate{\Omega_{02}} & \qw                               & \qw   & \qw          
}
}
\end{center}
\begin{center}
\mbox{
\Qcircuit @C=.8em @R = .5em {
& \multigate{2}{D} & \qw &     & & \qw & \gate{Z_0} & \qw         & \qw          & \ctrl{1}    & \ctrl{2}    & \qw  & \qw    \\
& \ghost{D}             & \qw & = & & \qw & \qw         & \gate{Z_1} & \qw          & \gate{\Phi_{01}} & \qw         & \ctrl{1} & \qw\\
& \ghost{D}             & \qw &    & & \qw & \qw         & \qw         & \gate{Z_2} & \qw          & \gate{\Phi_{02}} & \gate{\Phi_{12}}    & \qw      \;\;\; .
}
}
\end{center}
\end{small}
Here, the unitary operators are the two-qubit controlled-phase operators $\Omega_{ij}=\Omega_{ij}(\omega)$, where $\omega = \exp(2\pi i/8)$, the single-qubit operators 
\begin{eqnarray}
Z_0 & = & \exp{(-i \gamma c_0 \sigma^0_z \Delta t)} \nonumber \\
Z_1 & = & \exp{(-i \gamma c_1 \sigma^1_z \Delta t)} \nonumber \\
Z_2 & = & \exp{(-i \gamma c_2 \sigma^2_z \Delta t)} \nonumber
\end{eqnarray}
and the two-qubit controlled-phase operators 
\begin{eqnarray}
\Phi_{01} & = & \exp{(-i \gamma c_3 \mathrm{diag}(1,1,1,-1)_{01} \Delta t)} \nonumber \\
\Phi_{02} & = & \exp{(-i \gamma c_4 \mathrm{diag}(1,1,1,-1)_{02} \Delta t)} \nonumber \\
\Phi_{12} & = & \exp{(-i \gamma c_5 \mathrm{diag}(1,1,1,-1)_{12} \Delta t)} \; ,\nonumber
\end{eqnarray}
where
\begin{equation} \nonumber
  \gamma = -(2\pi/8)^2/\sqrt{8}
\end{equation}
and 
\begin{eqnarray}
c_0 & = & -1.42 \nonumber \\
c_1 & = & -5.66 \nonumber \\
c_2 & = & -22.63 \nonumber \\
c_3 & = & 22.63 \nonumber \\
c_4 & = & 11.31 \nonumber \\
c_5 & = & -5.66 \; . \nonumber
\end{eqnarray} 
Note that, in principle, a seventh three-qubit operator would also be necessary proportional to the vector $(1,1,1,1,1,1,1,-1)$ along the diagonal, but for the (bit-swapped) diagonal of the $8$ lattice point simulation, $(0,16,4,4,1,9,9,1)$, its coefficient is identically zero. This circuit requires a total of $19$ gates ($10$ single-qubit and $9$ two-qubit). 

In Fig. 2, we show results from a three-qubit simulation with a double-well potential, where each well is resolved with two lattice points. The time step $\Delta t = 1/5$ and $v = 5$. The initial state was $|\psi_{init}\rangle = |110\rangle$. Because the initial state only occupies half of one well, oscillatory dynamics are visible within the well. After a few time steps, the oscillatory state tunnels between wells. Oscillatory dynamics are evident within $4$ time steps, but $5$ time steps are required before the oscillatory state tunnels appreciably to the second well and $7$ time steps are needed to see oscillation of the tunneled state. Thus, between $4 \times 19 = 76$ or $7 \times 19 = 133$ gates would be required in order to see interesting tunneling effects in such a simulation. Four-qubit simulations can be envisaged using circuits based on similar methodology.

\begin{figure}
\epsfig{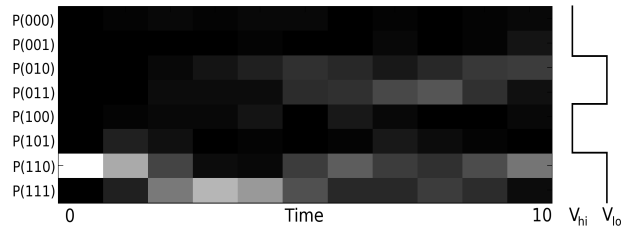}
\caption{Particle probability distribution as a function of time over ten time steps in a three-qubit double-well simulation. The potential schematic is shown to the right, as in Fig. 1. Gray scales are as in Fig. 1. The double-well potential has two lattice points per well and, for the given initial state, an oscillation is induced in one of the wells then tunnels to the other well.}
\label{threequbit}
\end{figure}

{\it Conclusion} Only $n = \log_2{N}$ qubits are required for an $N$ lattice-point particle simulation, therefore this algorithm is efficient in the number of qubit resources required. With a very few qubits, interesting tunneling dynamics may be simulated with a gate count that is within reach of current quantum architectures.

{\bf Acknowledgements} This work was supported by NSF grants PHY 0939853 and DMS 1029764. The author would like to thank Mike Geller and Phillip Stancil in the Physics Department at the University of Georgia for helpful comments and discussions.

\end{document}